\documentclass[12pt]{iopart}

\usepackage{graphicx}

\newcommand{\vect}[1]{\bi{#1}}

\begin{document}

\title{Speckles generated by skewed, short-coherence light beams}

\author{D. Brogioli$^1$ , D. Salerno$^1$ , F. Croccolo$^2$ , 
  R. Ziano$^1$ and F. Mantegazza$^1$}
\address{$^1$ Dipartimento di
  Medicina Sperimentale, Universit\`a degli Studi di Milano - Bicocca,
  Via Cadore 48, Monza (MI) 20052, Italy. \\
  $^2$ Physics Department,
  University of Fribourg, Ch. du Mus\'ee 3, 1700 Fribourg
  Switzerland.}


\begin{abstract}
  When a coherent laser beam impinges on a random sample (e.g. a
  colloidal suspension), the scattered light exhibits characteristic
  speckles.  If the temporal coherence of the light source is too
  short, then the speckles disappear, along with the possibility of
  performing homodyne or heterodyne scattering detection or photon
  correlation spectroscopy.  Here we investigate the scattering of a
  so-called ``skewed coherence beam'', i.e., a short-coherence beam
  modified such that the field is coherent within slabs that are
  skewed with respect to the wave fronts. We show that such a beam
  generates speckles and can be used for heterodyne scattering
  detection, despite its short temporal coherence.  Moreover, we show
  that the heterodyne signal is not affected by the multiple
  scattering. We suggest that the phenomenon presented here can be
  used as a mean to perform heterodyne scattering measurement with any
  short-coherence radiation, including X-rays.
\end{abstract}

\section{Introduction}

Observing of the light diffused from a scattering sample is a widely
used, well-known tool applied to the study of biophysical systems,
colloidal suspensions, and complex fluids
\cite{scheffold2007}. Homodyne dynamic light scattering and heterodyne
detection are based on interference and, in particular, on the
observation of speckle patterns. In this case, a long coherence is
needed because the light path difference of the interfering beams must
not exceed the coherence length. On the other hand, the disappearance
of interference beyond the coherence length may be exploited to select a
well-defined slab from a thick sample, for example in optical coherence
tomography \cite{huang1991}.

A single-mode laser beam has a very long longitudinal coherence and is
transversally coherent across its section. In a beam with a short
longitudinal coherence, the regions in which the electric field is
correlated are slabs perpendicular to the beam direction. Using a
dispersing optical element, it is possible to manipulate such a beam
so that the coherent slabs are skewed by an angle $\sigma$ with
respect to the plane perpendicular to the beam direction
\cite{picozzi2002}.  This effect has been extensively studied in the
context of ultra-short optical pulses \cite{martinez1986, porras2003},
and it can be utilized to achieve more efficient nonlinear pulse
generation (by increasing the phase-matching condition)
\cite{martinez1989, szabo1990}, or to avoid some linear side effects
(such as group velocity mismatch and walk-off) \cite{ditrapani1998}.

In this paper, we describe the scattering of a skewed coherence beam
by a random medium. In the near-field images of the sample, we observe
that a skewed coherence beam gives rise to a speckle pattern that is not
visible using short-coherence illumination, although the coherence
length is identical. Thanks to the visibility of the speckle pattern,
our experimental setup with the skewed coherence beam can be used as a
so called Scattering In Near Field (SINF) device \cite{brogioli2008,
  croccolo2011} that operates in a heterodyne detection configuration.
Accordingly, we are able to obtain heterodyne light scattering
measurements despite the short coherence of the illumination.

\section{Experimental setup}


\begin{figure}
\includegraphics{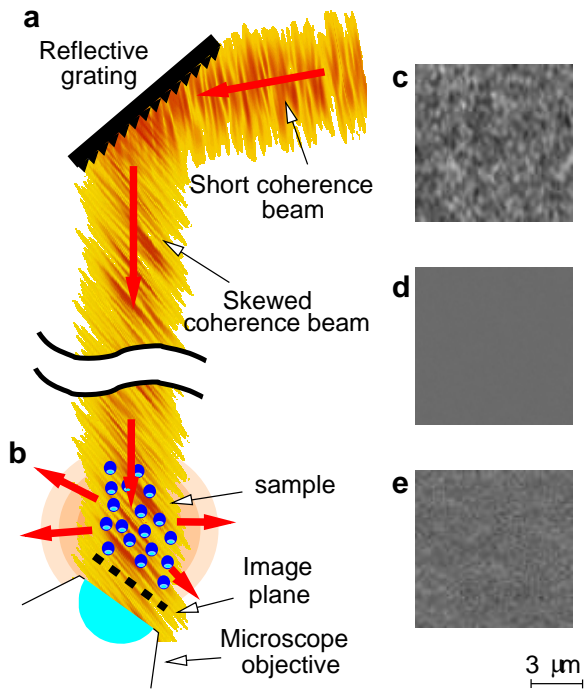}
\caption{
\label{fig:setup}
A schematic of the experimental setup. (a), generation of
the skewed beam.  The coherent slabs of the original beam are
perpendicular to the propagation direction. In the first order
diffracted beam the slabs are skewed with respect to the propagation
direction.  (b), the sample (colloidal suspension) and collection
optics. Typical near-field images obtained with various beams 
are shown: a laser beam (c), a short coherence beam (d), and
a skewed coherence beam with skew angle of $\sigma=47^{\circ}$ (e).
 The scattering medium is a water suspension of 80~nm
polystyrene nanospheres.  }
\end{figure}


Our short-coherence light source is a laser diode
driven under threshold (Sacher Lasertechnik SAL-0660-025).
The emission has a maximum at a wavelength of 660~nm, with an
8~nm FWHM bandwidth. The regions of the emitted beam in
which the field is coherent consist of thin slabs that are orthogonal
to the direction of propagation and whose thickness equals the
longitudinal coherence length (i.e. about 17~$\mu$m).


To obtain the desired skewed coherence, the beam is
diffracted by a reflective grating \cite{porras2003, picozzi2002},
with 600 lines per millimeter, blazed at $17.5^{\circ}$, and the first
order diffracted beam is selected (see Fig.~\ref{fig:setup}a). In the
resultant beam, the regions in which the field is coherent are slabs
that are skewed (inclined) by an angle $\sigma$ with respect to a direction
perpendicular to the propagation direction \cite{porras2003,
  picozzi2002}.  In our case, the skew angle $\sigma$ can be adjusted
from about $\sigma=20^{\circ}$ to $\sigma=50^{\circ}$ by changing the
incidence angle of the beam with respect to the grating.


As shown in Fig.~\ref{fig:setup}b,
the skewed beam passes through a 1~mm-thick cell containing 
the scattering sample (e.g. a colloidal suspension).

The intensity of the light in a plane close to 
the exit face of the cell is imaged by a CCD camera (Luca Andor) through
a high numerical aperture microscope objective (MO; Nikon CFI Plan
Apochromat 63X NA 1.4), which collects both the transmitted and 
scattered beams (heterodyne detection).

\section{Experimental results}


The images obtained by illuminating the sample with a laser beam
(Fig.~\ref{fig:setup}c) clearly show the usual near field speckle patterns 
\cite{goodman_speckles, giglio2001},
with a strong contrast, due to the random interference of the 
scattered beams. When a non-skewed
beam (i.e. $\sigma=0$) with a short temporal coherence is used
(Fig.~\ref{fig:setup}d), the speckles are almost invisible 
and appear extremely smeared because no interference can occur.
Quite surprisingly, the images obtained with the skewed beams 
(i.e. $\sigma>0$) do show speckles (Fig.~\ref{fig:setup}e), although
their texture is different from that which is observed for laser light.


\begin{figure}
\begin{center}
\begin{tabular}{cc}
(a), $\sigma = 0^{\circ}$ & (b), $\sigma =22^{\circ}$\\
\includegraphics{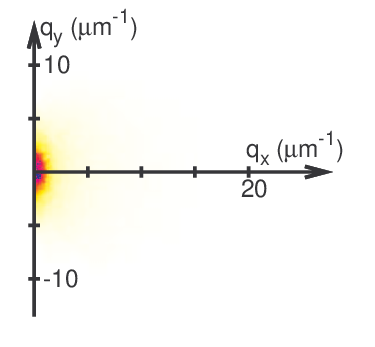} &
\includegraphics{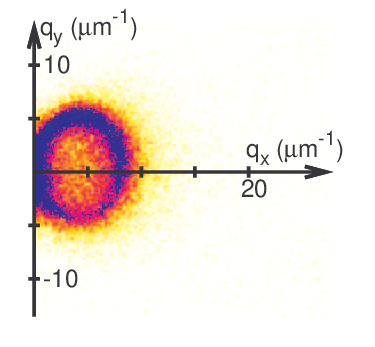} 
\\
(c), $\sigma = 34^{\circ}$ & (d), 
$\sigma = 47^{\circ}$\\ 
\includegraphics{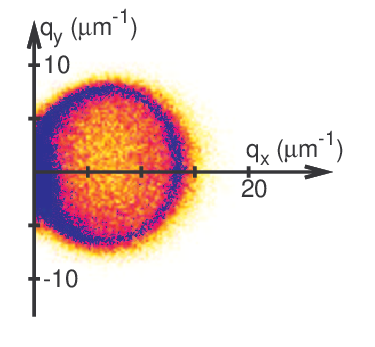} &
\includegraphics{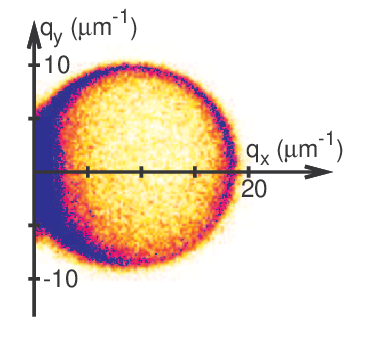}\\
\multicolumn{2}{c}{(e) 
\includegraphics{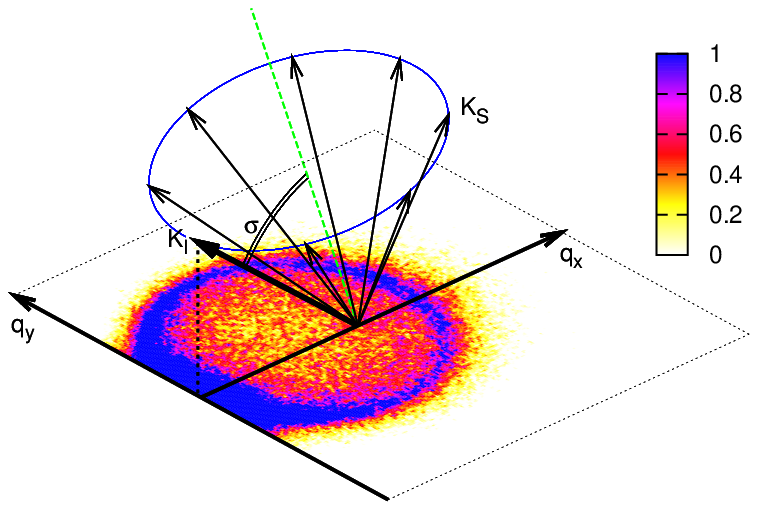}
}
\end{tabular}
\caption{
\label{fig:spectra}
Power spectra.  (a), (b), (c), (d),
Two-dimensional false-color power spectra of the near-field speckle
patterns of the images obtained with an impinging beam with a short
coherence, at various skew angles $\sigma$. (a), the spectrum of
the image in Fig.~\ref{fig:setup}d, using a non-skewed beam. (b),
(c), spectra of the images at a skew angle $\sigma=22^{\circ}$ and
$\sigma=34^{\circ}$. (d), the spectrum of the image in
Fig.~\ref{fig:setup}e, using a skew angle $\sigma=47^{\circ}$.  The
scattering medium is a water suspension of 80~nm polystyrene
particles.  (e), A three dimensional schematic of the
relationship among the wave vector $\vect{K_I}$ of the impinging beam,
the wave vectors $\vect{K_S}$ of the scattered beams, and the wave
vectors $\vect{q}=(q_x,q_y)$ of the interference fringes in the image
plane.  The most intense points of the power spectrum are generated by
light scattered along a cone, whose axis (green dashed line) is
perpendicular to the coherent slabs.  }
\end{center}
\end{figure}

To characterize the speckle fields, we evaluated the power 
spectra of the images obtained with beams skewed at various angles;
the results are reported in Fig.~\ref{fig:spectra}. 
The power spectrum at $\sigma=0$ exhibits a weak signal for the wave
vectors $\vect{q}$ close to the center of the Fourier-space image:
this corresponds to the faint speckles. 
As the skew angle $\sigma$ is increased, higher frequency modes 
appear in the images; in the power spectra, they appear as ellipses.
The size of the ellipse in the power spectrum increases as the skew
angle $\sigma$ is increased.


These images can be interpreted as holograms of the scattered field.
Each point in the near-field image (with wave vector $\vect{q}$) 
is generated by the interference between the most intense transmitted
beam (with wave vector $\vect{K_I}$) and a single scattered beam (with
wavevector $\vect{K_S}$). The relationship among $\vect{K_I}$,
$\vect{K_S}$ and $\vect{q}$ is shown graphically in
Fig.~\ref{fig:spectra}e; the wave vector $\vect{q}$ is the projection
of the transferred wave vector $\vect{Q}=\vect{K_S}-\vect{K_I}$ in the
image plane \cite{brogioli2009bis}.  The SINF technique 
\cite{brogioli2008, brogioli2009, brogioli2009bis, cerbino2008} exploits
this relationship to measure the scattering intensity by acquiring near-field
images. In this case, the following observations can be made:
{\em i}) the observed ellipses correspond to scattering along a cone;
{\em ii}) the impinging beam direction is a generatrix of the cone;
{\em iii}) the axis of the cone is orthogonal to the coherent slabs.
In reality, the
sample scatters at all angles, but only the scattering along the
cone beats coherently with the local oscillator to provide a
detectable heterodyne signal. Thus it will be called the ``scattering
detection cone'' (SDC).


\begin{figure}
\begin{center}
\includegraphics{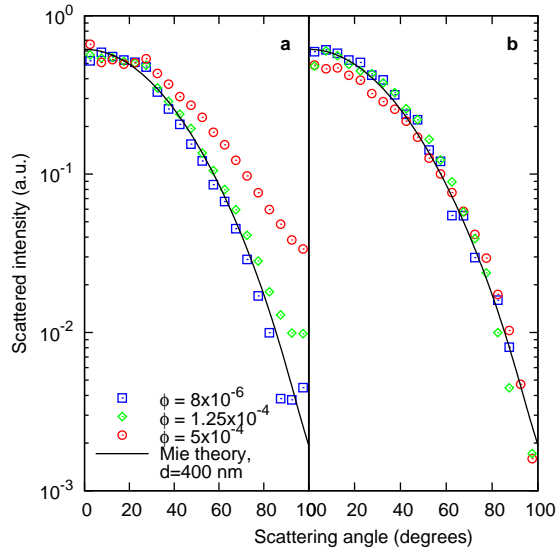}
\caption{
\label{fig:scattering}
Measured scattering intensity. The sample consists of 400~nm polystyrene 
particles in water at various volume fractions $\phi$. The data were obtained
using a laser beam (a) or a skewed $\sigma=47^{\circ}$
beam (b). The continuous line represents the Mie theory prediction 
for 400~nm dielectric spheres.
}
\end{center}
\end{figure}

Following the idea of SINF, we can recover the scattering intensities
from the power spectra. The results for both a laser beam and a skewed
coherence beam are shown in Fig.~\ref{fig:scattering}. In both cases,
at a low volume fraction, the measured
scattering intensity closely follows the Mie theory
\cite{vandehulst}. This shows that the detected light, scattered along
the SDC, has the same direction and intensity as the light scattered
by the particles, and the SDC acts only as a mask.  By skewing the
coherence, we can perform heterodyne detection, despite the
short coherence of our light source.

When using laser light, we observe a progressive departure from
the Mie theory for large scattering wave
vectors \cite{vandehulst}, that can be easily interpreted to be due to
multiple scattering. In contrast, the use of the skewed beam 
(Fig.~\ref{fig:scattering}b) results in an almost perfect overlap
between the data and the theoretical results single scattering,
independent of the nanoparticle concentration. This experiment shows that the use
of skewed coherence beams in a heterodyne scattering detection setup 
suppresses the detection of multiple scattered light.

\section{Discussion}

\begin{figure}
\begin{center}
\includegraphics{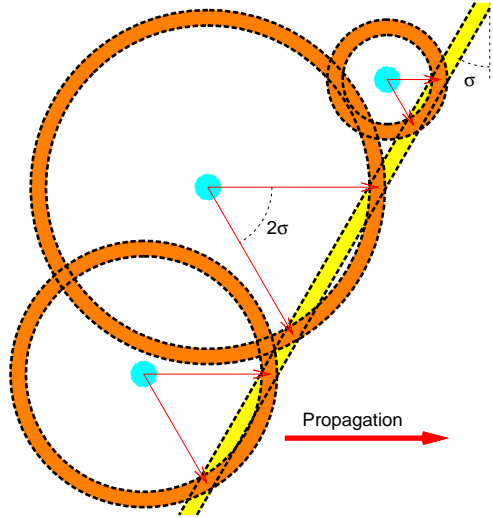}
\end{center}
\caption{
\label{fig:explanation}
A qualitative interpretation of the scattering mechanism in the presence
of a skewed beam.
The impinging beam propagates from left to right
in the direction shown by the arrow. The skewed coherent slab is shown
as a yellow stripe. When it hits the particles (blue dots),
secondary spherical waves are generated. The parts of the 
secondary wave that are coherent with the slab are the spherical shells
(orange rings). Heterodyne interference is only observed where
the coherent slab and the coherent spherical shells overlap.}
\end{figure}

Figure~\ref{fig:explanation} presents a schematic explanation of the
obtained results. A coherent slab impinges on a set of particles,
which emit secondary waves. Interference only occurs where 
there is overlap between the coherent slab and
the parts of the secondary waves that are coherent with it. In the 
section shown, overlap only occurs at two points, which define
the two directions (indicated by arrows) along which the scattered
waves interfere with the impinging beam. One is always along the forward
direction, and the other is at a scattering angle of $2\sigma$. This
direction is identical for every particle.  In three dimensions, the
intersection of the coherent slab with the secondary spherical wave
occurs along a circle, which defines a scattering cone with
properties identical to those of the SDC.

From this simplified picture of the interference generated by
a skewed beam, one can also infer why multiple scattering is not detected
in this configuration. In fact, light that has been scattered more than
once will arrive at
the screen plane with a time shift greater than the beam's temporal
coherence, thereby preventing interference.

When we take into account the thickness of the coherent slabs, we
notice that the secondary waves indeed interfere with the impinging
beam for an interval of directions around $\vartheta=2\sigma$, whose
extent $\delta \vartheta$ increases as the distance from the coherent
slab of the impinging beam to the emitting particle decreases. In
particular, if we analyze a plane inside the sample, the particles
whose distance from the detection plane is less than the coherence
length can give an heterodyne signal at every angle. In other words,
only a thin volume around the detection plane contributes to the
scattering signal at a generic angle, but the whole volume of the
sample contributes to the heterodyne signal along the SDC.  This
explains why the heterodyne signal along the SDC is much higher than
along other directions.

In general, the manipulation of the coherent slab allows to select the
volume of the sample which is detected: in our experiments, we are
interested in extending this volume up to the whole sample; on the
other hand, the so-called ``optical coherence tomography''
\cite{huang1991} exploits the short coherence for selecting a well
defined slice of the sample (this is the meaning of ``tomography'',
namely ``drawing the image of a slice''). This view makes evident the
deep analogy between the rejection of the multiple scattering in our
experiment and the rejection of the slices other than the selected one
in optical coherence tomography.

\section{Applications}


Currently, measuring the skew angle of a sample beam is performed
using non-linear optics and only in the case of ultra-short pulses
with high energy \cite{tipa}. However, the phenomenon described here
enables the development of a much simpler technique based on observing
the SDC in the power spectra; the axis of the SDC directly provides the normal to
the coherence slabs. Such a technique would also be effective for 
faint beams, and both for continuous waves and ultra-short pulses.

Furthermore, multiple scattering is a serious issue affecting
scattering measurements, and several techniques have been developed to
suppress the contributions of multiple scattering
(e.g. ref.~\cite{moitzi2009, pusey1999} and references therein).
This experiment shows that the use of skewed coherence beams in
a heterodyne scattering detection setup suppresses multiple
scattering and represents the most natural and effective way
to reach this target.

In the case of X-rays \cite{sutton2008, sutton1991, mochrie1997,
  cerbino2008bis, nugent2010} the coherence conditions necessary for
obtaining interference can be achieved by synchrotron radiation or
FEL. However, this only occurs after filtering through a
monochromator, which strongly reduces the photon flux. Our findings
suggest that skewing the coherence of a short-coherence X-ray beam can
enable heterodyne detection. 

The actual optical scheme will depend on the application. For example,
the main beam of a small-angle X-ray scattering (SAXS) apparatus can
be used. In this case, the angular dispersion of the beam is much less
than 1/1000 of radiant, which ensures a transversal coherence length
of the order of the micron, while the transversal coherence is of the
order of a few nanometers. In order to obtain the skewed-coherence
beam, a transversally coherent region of the main beam is selected by
means of a pin-hole, the resulting beam is sent through a transmission
grating with a line spacing of the order of 100~nm, and the
first-order diffracted beam is selected by a second pin-hole.  The
resulting beam has a skew angle of the order of the milliradiants, and
the corresponding SDC has an aperture of the same order, which is in
the range usually detected with SAXS. The SINF detection scheme can
still be applied with a variant with respect to the above-described
method. Instead of taking an image of a plane, we will place an
intensity mask (a transmission grating) on the same plane, with the
wave vector we want to measure. The transmitted intensity will
represent the amplitude of the corresponding Fourier mode of the image,
and will represent the heterodyne signal.

\section{Conclusions}

We have shown that the use of a skewed-coherence beam allows to detect
a heterodyne signal also with short-coherence light. We show that it
is possible to skew the coherence of a short-coherence beam by means
of an optical system including a diffraction grating. Anyhow, the
skewing optical system do not increase the coherence length, nor it
acts as a narrow bandwidth filter. The detection is possible only for
light scattered along the SDC, while the other scattering directions
give a negligible heterodyne signal. The axis of the SDC is
perpendicular to the coherent slabs, and thus the detection of the SDC
represents a very effective method for measuring the coherence
skewness of either a continuous wave or a pulsed beam.  When applied
to quite turbid samples, the technique has the remarkable advantage of
suppressing the multiple scattering contribution of the scattering
signal. We suggest that the phenomenon presented here can be used as a
mean to perform heterodyne scattering measurement with any
short-coherence radiation, and the application of the technique to
X-rays has been discussed.

\section*{Bibliography}

\bibliography{coherence}

\section*{Acknowledgements}
This work was partially financially supported by the EU (project
NAD CP-IP 212043-2). F.C. acknowledges
his present support from the European Union (Marie Curie funding,
contract IEF-251131; DyNeFl project).

\end{document}